\documentclass[12pt]{article}
\usepackage{amsmath}
\usepackage{amssymb}
\usepackage{graphicx}
\usepackage{bm}
\usepackage{upgreek}

\begin{document}

\begin{center}
{\bf \Large Quasi-Rip and Pseudo-Rip universes induced by the fluid inhomogeneous equation of state }

\bigskip

I. Brevik\footnote{iver.h.brevik@ntnu.no}

\bigskip
Department of Energy and Process Engineering, Norwegian University of Science and Technology, N-7491 Trondhein, Norway

\bigskip
V. V. Obukhov and A. V. Timoshkin

\bigskip

Tomsk State Pedagogical University, 634061 Tomsk, Russia

\end{center}

\begin{abstract}

We investigate specific models for a dark energy universe leading to Quasi-Rip and Pseudo-Rip cosmologies. In the Quasi-Rip model the equation of state parameter $w$ is less than $-1$ in the first stage, but becomes larger than $-1$ in the second stage. In the Pseudo-Rip model the Hubble parameter tends to a constant value in the remote future, although $w$ is always less than $-1$. Conditions for the appearance of the Quasi-Rip and the Pseudo-Rip in terms of the parameters in the equation of state are determined. Analogies with the theory of viscous cosmology are discussed.

\end{abstract}

\bigskip

\section{Introduction}

The discovery of the accelerated expansion of the universe led to the appearance of new theoretical models in cosmology (for recent reviews, see \cite{bamba12,nojiri11}), and significantly changed our understanding of its evolution. Recent observations indicate that the universe is dominated by a negative-pressure dark energy component (a dark fluid). Quintessence/phantom dark energy proposed to explain the cosmic acceleration should imply a strong negative pressure (positive tensile stress). The fluid can be characterized by an equation of state parameter  $w=p/\rho$ being less than $-1$, where $p$ is the pressure and $\rho$ the energy density. The condition $w<-1$ corresponds to a dark energy density that increases monotonically with time $t$ with scale factor $a$. There are several interesting possible scenarios concerning the fate of the universe, including Big Rip \cite{caldwell03,nojiri03}, Little Rip \cite{frampton11,brevik11,frampton12,astashenok12,astashenok12B,astashenok12C,nojiri11B,makarenko12,brevik12C}, and Pseudo-Rip \cite{frampton12B} models. These models are based on the assumption that  the dark energy density $\rho$ is a monotonically increasing function.

In the present paper we are interested to study a cosmological model in which the dark energy density $\rho$ monotonically increases $(w<-1)$ in the first stage, and thereafter monotonically decreases $(w>-1)$. At first, it thus tends to disintegrate bound structures in the universe, but then in the second stage the disintegration becomes reversed, implying that already disintegrated structures have the possibility to be recombined again. This cosmological scenario is called Quasi-Rip \cite{wei12}. Another interesting possibility for the evolution of the universe is the so-called Pseudo-Rip, where the Hubble parameter, although increasing, tends to a "cosmological constant" in the remote future. That means, $H(t) \rightarrow H_\infty <\infty, t\rightarrow +\infty$, where $H_\infty$ is a constant.

We shall examine the influence from the equation of state for the dark fluid, explicitly dependent on $w$ as well as the cosmological constant $\Lambda$, upon the occurrence of the Quasi-Rip and  the Pseudo-Rip. In the final section we discuss how the Quasi-Rip phenomenon may be interpreted in terms of a bulk viscosity in the dark fluid.

\section{Dark fluid inhomogeneous equation of state in the Quasi-Rip model}

As an explicit model for the Quasi-Rip, let us choose the energy density $\rho$ to be a function of the scale factor $a$. A simple form is \cite{wei12}
\begin{equation}
\rho=\rho_0 a^{\alpha-\beta \ln a}, \label{1}
\end{equation}
where $\alpha$ and $\beta$ are constants and $\rho_0$ the density at a specified time $t_0$.

The derivative of $\rho$ with respect to cosmic time $t$ is
\begin{equation}
\dot{\rho}=\rho H(\alpha-2\beta \ln a), \label{2}
\end{equation}
where $H=\dot{a}/a$ is the Hubble rate.

Assume now that the universe is filled with an ideal fluid (dark energy) obeying an inhomogeneous equation of state \cite{nojiri05},
\begin{equation}
p=w(a)\rho+\Lambda(a), \label{3}
\end{equation}
where $w(a)$ and $\Lambda(a)$ depend on the scale factor $a$.

The energy conservation law is
\begin{equation}
\dot{\rho} +3H(\rho+p)=0. \label{4}
\end{equation}
Taking into account Eqs.~(\ref{2})-(\ref{4}) we obtain
\begin{equation}
\rho(\alpha -2\beta \ln a)+\rho(1+w(a))+\Lambda (a)=0. \label{5}
\end{equation}
Let us suppose that the "cosmological constant" $\Lambda(a)$ is proportional to the energy density, i.e.
\begin{equation}
\Lambda(a)=\gamma \rho_0^2a^{2(\alpha -\beta \ln a)}, \label{6}
\end{equation}
where $\gamma$ is some constant.

From (\ref{5}) and (\ref{6}) we obtain
\begin{equation}
w(a)=-1-\gamma \rho_0 a^{\alpha-\beta \ln a} +2\beta \ln a -\alpha. \label{7}
\end{equation}
If we require $\beta>0$, then the extremum of $\rho$ is a maximum. It occurs when $a=e^{\frac{\alpha}{2\beta}}$.Then the equation of state parameter is $w(a)=-1-\gamma \rho_0 a^{\alpha/2}$.

Consequently, if we assume an ideal fluid which obey an equation of state (\ref{3}) and (\ref{5}), then we obtain a solution realizing the Quasi-Rip (\ref{1}). Note that such a Quasi-Rip is exponential, caused by the cosmological "constant" $\Lambda$.

Let us solve equation (\ref{5}) with respect to $\Lambda(a)$, when the parameter $w(a)$ is chosen as
\begin{equation}
w(a)=-1-\frac{\delta}{3\rho_0}a^{\beta \ln a-\alpha}, \label{8}
\end{equation}
where $\delta$ is a constant. The result becomes
\begin{equation}
\Lambda(a)=-\rho_0a^{\alpha -\beta \ln a}(\alpha-2\beta \ln a)-\frac{\delta}{3}. \label{9}
\end{equation}
In this case the Quasi-Rip is caused by the quantity $w$. The future behavior of our universe will be dependent on the choice of the model parameters $\alpha$ and $\beta$.

Thus, we have explored the equation of state (\ref{3}), yielding the Quasi-Rip.

\section{Phantom energy models with asymptotically de Sitter evolution}

Now we investigate a class of models with monotonically increasing dark energy density for which the expansion of the universe asymptotically approaches the exponential regime. The Hubble rate tends to a constant value (cosmological constant or asymptotically de Sitter space). It may correspond to a Pseudo-Rip model.

Let us assume the following expression for the pressure of the dark energy:
\begin{equation}
p=-\rho-f(\rho), \label{10}
\end{equation}
where $f(\rho)$ is a function of the dark energy density $\rho$. From the Friedmann equation and the conservation law for a spatially flat universe the following relation results between the time $t$ and $f(\rho)$:
\begin{equation}
t=\frac{2}{\sqrt 3}\int_{x_0}^x \frac{dx}{f(x)}, \quad x \equiv \sqrt{\rho}, \label{11}
\end{equation}
where $x_0=\sqrt{\rho_0}$, $\rho_0$ being the energy density at present.

The expression for the scale factor has the following form:
\begin{equation}
a=a_0\exp \left( \frac{2}{3}\int_{x_0}^x \frac{xdx}{f(x)}\right). \label{12}
\end{equation}
For instance, let us assume \cite{astashenok12} that
\begin{equation}
f(x)=A\left( 1-\frac{x}{x_f}\right)~\alpha, \label{13}
\end{equation}
where $A$ and $\alpha$ are positive constants. We assume $\alpha \geq 1$. In this case the integral (\ref{11}) diverges at some finite value $x=x_f <\infty$.

If $\alpha \neq 1,2$ one obtains the following expression for the scale factor:
\[
a=a_0\exp \left( \frac{x_f t}{\sqrt 3}\right) \exp [g_\alpha (t)], \]
\begin{equation}
g_\alpha (t)=\frac{2x_f^2}{3A(2-\alpha)}\left[ \frac{\sqrt{3}A(\alpha-1)t}{2x_f}+\left(1-\frac{x_0}{x_f}\right)^{1-\alpha}\right]^{1+\frac{1}{1-\alpha}}. \label{14}
\end{equation}
The Hubble parameter becomes
\begin{equation}
H=\frac{x_f}{\sqrt 3}\left\{ 1+\frac{\alpha-1}{2-\alpha}\left[ \frac{\sqrt{3}(\alpha -1)At}{2x_f}+\left(1-\frac{x_0}{x_f}\right)^{1-\alpha}\right]^{\frac{\alpha}{1-\alpha}}\right\}. \label{15}
\end{equation}
When $t\rightarrow \infty$, then the Hubble parameter $H\rightarrow x_f/\sqrt 3$. Therefore the expression (\ref{15}) asymptotically tends to the de Sitter solution.

The derivative of the energy density with respect to the cosmic time is
\begin{equation}
\dot{\rho}=\frac{3A}{k^2}\frac{\alpha-1}{\alpha-2}\left[\frac{\sqrt{3}(\alpha-1)At}{2x_f}+
\left(1-\frac{x_0}{x_f}\right)^{1-\alpha}\right]^{\frac{\alpha}{1-\alpha}} H. \label{16}
\end{equation}
From (\ref{3}), (\ref{4}), (\ref{15}) and (\ref{16}) we obtain the energy conservation law in the following form:
\begin{equation}
\frac{A}{k^2}\frac{\alpha-1}{\alpha-2}\left[\frac{\sqrt{3}(\alpha-1)At}{2x_f}+\left(1-\frac{x_0}{x_f}\right)^{1-\alpha}\right]^{\frac{\alpha}{1-\alpha}}+\frac{3}{k^2}[1+w(t)]H^2+\Lambda (t)=0, \label{17}
\end{equation}
where $w(t)$ and $\Lambda(t)$ are time dependent parameters.

Solving equation (\ref{17}) with respect to $w(t)$ we have
\begin{equation}
w(t)=-1-\frac{k^2\Lambda(t)}{3H^2}   -\frac{A}{3H^2}\frac{\alpha-1}{\alpha-2}\left[\frac{\sqrt{3}(\alpha-1)At}{2x_f}+\left( 1-\frac{x_0}{x_f}\right)^{1-\alpha}\right]^{\frac{\alpha}{1-\alpha}}. \label{18}
\end{equation}
Let us suppose that the parameter $\Lambda(t)$ is proportional to the square of the Hubble parameter \cite{houndjo11},
\begin{equation}
\Lambda(t)=\gamma H^2, \label{19}
\end{equation}
where $\gamma$ is a positive constant. Taking (\ref{19}) into account we obtain
\begin{equation}
w(t)=-1-\frac{k^2\gamma}{3}-\frac{A}{3H^2}\frac{\alpha-1}{\alpha-2}\left[ \frac{\sqrt{3}(\alpha-1)At}{2x_f}+\left( 1-\frac{x_0}{x_f}\right)^{1-\alpha}\right]^{\frac{\alpha}{1-\alpha}}. \label{20}
\end{equation}
If $t\rightarrow \infty$, $w(t)\rightarrow -1-k^2\gamma/3 <-1$. It corresponds to a dark fluid.

Now writing the parameter $w(t)$ in the form
\begin{equation}
w(t)=-1-\frac{\delta}{H^2}, \label{21}
\end{equation}
with $\delta$ a positive constant, we obtain from (\ref{17})
\begin{equation}
\Lambda(t)=\frac{3\delta}{k^2}-\frac{A}{k^2}\frac{\alpha-1}{\alpha-2}\left[ \frac{\sqrt{3}(\alpha-1)At}{2x_f}+\left( 1-\frac{x_0}{x_f}\right)^{1-\alpha}\right]^{\frac{\alpha}{1-\alpha}}. \label{22}
\end{equation}
Of main interest are the particular cases when $\alpha=1$ or $\alpha=2$.

Let us first put $\alpha=1$. We will investigate this kind of model, following from (\ref{13}), in analogy with the earlier model (\ref{14}).

The scale factor becomes

\[ a(t)=a_0\exp \left( x_f\frac{t}{\sqrt 3}\right)\exp [g_1(t)], \]
\begin{equation}
g_1(t)=\frac{2x_f^2}{3A}\left( 1-\frac{x_0}{x_f}\right) \left( \exp \left( -\frac{\sqrt{3}At}{2x_f}\right)-1\right). \label{23}
\end{equation}
We calculate the Hubble parameter
\begin{equation}
H=\frac{x_f}{\sqrt 3}\left[ 1-\left(1-\frac{x_0}{x_f}\right)\exp \left(-\frac{\sqrt{3}At}{2x_f}\right) \right], \label{24}
\end{equation}
and the time derivative of the energy density,
\begin{equation}
\dot{\rho}=\frac{3A}{k^2}\left( 1-\frac{x_0}{x_f}\right) \exp \left(-\frac{\sqrt{3}At}{2x_f}\right) H. \label{25}
\end{equation}
Taking into account (\ref{3}), (\ref{4}), (\ref{24}) and (\ref{25}) we can rewrite the energy conservation equation as
\begin{equation}
\frac{A}{k^2}\left(1-\frac{x_0}{x_f}\right) \exp \left( -\frac{\sqrt{3}At}{2x_f}\right)+\frac{3}{k^2}[1+w(t)] H^2+\Lambda (t)=0. \label{26}
\end{equation}
Using (\ref{19}) to solve equation (\ref{26}) with respect to $w(t)$ we find
\begin{equation}
w(t)=-1-\frac{k^2\gamma}{3}-\frac{A}{3H^2}\left(1-\frac{x_0}{x_f}\right)\exp \left(-\frac{\sqrt{3}At}{2x_f}\right). \label{27}
\end{equation}
This shows that the Pseudo-Rip is connected with the Hubble parameter (\ref{24}). Solving $\Lambda(t)$ from (\ref{26}) and inserting $w(t)$ from (\ref{21}) we get
\begin{equation}
\Lambda(t)=\frac{3\delta}{k^2}-\frac{A}{k^2}\left( 1-\frac{x_0}{x_f}\right)\exp \left(-\frac{\sqrt{3}At}{2x_f}\right). \label{28}
\end{equation}

Now go on to consider the case $\alpha=2$.  The scale factor is
\[ a(t)=a_0 \exp \left( x_f\frac{t}{\sqrt 3 }\right)\exp [g_2(t)], \]
\begin{equation}
g_2(t)=\frac{2x_f^2}{3A}\ln \left[ \frac{\sqrt{3}A(x_f-x_0)}{x_f^2} t+1\right], \label{29}
\end{equation}
and the Hubble parameter is
\begin{equation}
H=\frac{x_f}{\sqrt 3}\left( 1+\frac{2x_f}{\sqrt{3}At+\frac{x_f^2}{x_f-x_0}}\right). \label{30}
\end{equation}
We now find
\begin{equation}
\dot{\rho}=-\frac{12Ax_f^2}{k^2}\frac{H}{\left(\sqrt{3}At+\frac{x_f^2}{x_f-x_0}\right)^2}, \label{31}
\end{equation}
and in view of (\ref{30}) and (\ref{31}), equation (\ref{4}) becomes
\begin{equation}
\frac{4Ax_f^2}{\left( \sqrt{3}At+\frac{x_f^2}{x_f-x_0}\right)^2}+3[1+w(t)]H^2+k^2\Lambda(t)=0. \label{32}
\end{equation}
Using (\ref{19}) to solve (\ref{32}) with respect to $w(t)$, we find
\begin{equation}
w(t)=-1-\frac{k^2\gamma}{3}+\frac{4Ax_f^2}{3H^2\left( \sqrt{3}At+\frac{x_f^2}{x_f-x_0}\right)^2}, \label{33}
\end{equation}
which shows that the Pseudo-Rip behavior is determined by the Hubble parameter (\ref{30}).

Solving $\Lambda(t)$ from (\ref{32}), and taking into account (\ref{21}), we obtain
\begin{equation}
\Lambda(t)=-\frac{3\delta}{k^2}+\frac{4Ax_f^2}{k^2\left(\sqrt{3}At+\frac{x_f^2}{x_f-x_0}\right)^2}. \label{34}
\end{equation}
Thus, we have demonstrated the occurrence of a Pseudo-Rip, based upon the equation of state (\ref{4}).

\section{Quasi-Rip as an effect induced by a  bulk viscosity}

 As we have seen, the main characteristic of the Quasi-Rip is that the dark energy density $\rho$first increases monotonically with time and thereafter, in the second stage, decreases monotonically. We now wish to point out that this kind of behavior can actually be interpreted as an effect coming from a bulk viscosity in the universe (assuming spatial isotropy, the effect of the shear viscosity is omitted). The condition is that there occurs some kind of phase transition in the future, connected, for instance, with the transition to a turbulent era. It may be of interest to point this analogy out, not least so because it demonstrates the large flexibility of the cosmological theory in general. In the following we give a brief account of this kind of description, following the recent treatise in \cite{brevik12}, Sect. VI; cf. also the review article \cite{brevik12B}.

Assume then that the dark fluid is a one-component  fluid, starting from the present time $t_0=0$ as
an ordinary viscous fluid with a bulk viscosity  $\zeta$, and develops
according to the Friedmann equations. We assume that in this first epoch
 $w<-1$,
meaning that the universe develops  towards a future
singularity.
Before this happens, however, at some instant which we shall call $t=t_*$, we imagine that there occurs
a  phase transition of the  fluid into a different state, for definiteness called a turbulent state, after which $w_\mathrm{turb} >-1$. The equation of state is in this epoch $p_\mathrm{turb}=w_\mathrm{turb}\rho_{\rm turb}$. For simplicity we take
$\zeta$, as well as $w(t<t_*)$ and $w_\mathrm{turb}(t>t_*)$, to be constants.
One may ask: What is the time evolution of the fluid, according to this model?

The problem can easily be solved, making use of the condition that the density
of the fluid has to be continuous at $t=t_*$. One gets in the viscous epoch  \cite{brevik05}, \cite{brevik10},
\begin{equation}
H=\frac{H_0\,e^{t/t_c}}{1-\frac{3}{2}|1+w|H_0t_c(e^{t/t_c}-1)}\, ,
\label{35}
\end{equation}
\begin{equation}
a=\frac{a_0}{\left[1-\frac{3}{2}|1+w|H_0t_c(e^{t/t_c}-1)
\right]^{2/(3|1+w|)}}\, ,
\label{36}
\end{equation}
\begin{equation}
\rho=\frac{\rho_0\, e^{2t/t_c}}{\left[1-\frac{3}{2}|1+w|H_0t_c(e^{t/t_c}-1)
\right]^2}\, ,
\label{37}
\end{equation}
where  $t_c$ is the `viscosity time'
\begin{equation}
t_c=(12\pi G \zeta)^{-1}\, .
\label{38}
\end{equation}
The values of $H_*, a_*, \rho_*$ at $t=t_*$ are thereby known.

In the turbulent epoch $t>t_*$ we can make use of the same expressions
(\ref{35}) - (\ref{37}) as above, only with substitutions
$t_c \rightarrow \infty~(\zeta \rightarrow 0)$, $t \rightarrow t-t_*,~ w
\rightarrow w_\mathrm{turb}$,
$H_0 \rightarrow H_*$, $a_0 \rightarrow a_*$, $\rho_0 \rightarrow \rho_*$. Thus
\begin{equation}
H=\frac{H_*}{1+\frac{3}{2}(1+w_\mathrm{turb})H_*(t-t_*)}\, ,
\label{39}
\end{equation}
\begin{equation}
a=\frac{a_*}{\left[ 1+\frac{3}{2}(1+w_\mathrm{turb})H_*(t-t_*)
\right]^{2/(3(1+w_\mathrm{turb}))}}\, ,
\label{40}
\end{equation}
\begin{equation}
\rho=\frac{\rho_*}{\left[ 1+\frac{3}{2}(1+w_\mathrm{turb})H_*(t-t_*)
\right]^{2}}
\label{41}
\end{equation}
(recall that $w_{\rm turb}>-1$). Thus the density $\rho$, at first increasing with increasing $t$ according to
Eq.~(\ref{37}),
decreases again once the turbulent era has been entered, and goes smoothly
to zero as $t^{-2}$ when $t\rightarrow \infty$.
In this way the transition to turbulence protects the universe from entering
the future singularity. We see again the essence of the Quasi-Rip phenomenon, now interpreted in terms of a bulk viscosity. In view of the phase transition the universe is protected from running into the future singularity, and may thus avoid the disintegration of bound structures.

It should be noted that whereas the density is continuous at $t=t_*$ the
pressure is not:
In the laminar era $p_*= w \rho_* <0$, while in the turbulent era $p_*
=w_\mathrm{turb}\,\rho_* $
will even  be positive if $w_\mathrm{turb}>0$.

\section{Conclusions}

The Quasi-Rip model has a unique property making it different from Big Rip, Little Rip, and Pseudo-Rip. All the last-mentioned models arise from the assumption that the dark energy density $\rho$ is monotonically increasing. This leads to the dissolution of all bound structures in the far future. Distinct from these models, in the Quasi-Rip model this assumption is broken. Our universe has according to this the possibility to rebuild after the rip. In the present work we have modeled the Quasi-Rip universe induced by the dark fluid inhomogeneous equation of state. We showed that Quasi-Rip cosmology can be caused exponentially  via the cosmological constant, or the parameter $w$. It would be of interest to understand if Quasi-Rip cosmology may be mapped with dark energy fluid cosmology mimicking string-landscape features \cite{elizalde12}.  From another side, the role of viscosity in Rip cosmology may be relevant also in the Quasi-Rip picture \cite{brevik12}.

We presented above also a dark energy model with an inhomogeneous equation of state, in which the Pseudo-Rip behavior is encountered in the far future.

 Finally, the viscous cosmology discussed in the last section ought  to be borne in mind, as a demonstration of the large versatility of the cosmological formalism especially as regards the later stages of the universe's evolution.

\bigskip

{\bf Acknowledgement}

\bigskip
We are grateful to Professor Sergei Odintsov for helpful discussions.

\end{document}